\documentclass[usegraphicx,usenatbib]{mn2e}
\bibpunct{(}{)}{;}{a}{}{,}
\usepackage{aasmacros}
\usepackage{color}

\newcommand{\kms}{\mbox{\,km\,s$^{-1}$}}

\newcommand{\eg}{e.g.,\ }
\newcommand{\ie}{i.e.,\ }
\newcommand{\Msun}{M_{\odot}}

\newcommand{\vph}{$v_{\rm ph}$}

\newcommand{\OI}{O~{\sc i}}

\newcommand{\CII}{C~{\sc ii}}
\newcommand{\NaI}{Na~{\sc i}}

\newcommand{\SiII}{Si~{\sc ii}}
\newcommand{\SiIII}{Si~{\sc iii}}
\newcommand{\SII}{S~{\sc ii}}
\newcommand{\SIII}{S~{\sc iii}}
\newcommand{\CaII}{Ca~{\sc ii}}

\newcommand{\FeII}{Fe~{\sc ii}}
\newcommand{\FeIII}{Fe~{\sc iii}}
\newcommand{\CoII}{Co~{\sc ii}}

\newcommand{\NiII}{Ni~{\sc ii}}

\newcommand{\Fefs}{$^{56}$Fe}
\newcommand{\Cofs}{$^{56}$Co}
\newcommand{\Nifs}{$^{56}$Ni}

\newcommand{\texp}{t_{\rm exp}}
\def\lsim{\mathrel{\rlap{\lower 4pt \hbox{\hskip 1pt $\sim$}}\raise 1pt\hbox {$<$}}}
\def\gsim{\mathrel{\rlap{\lower 4pt \hbox{\hskip 1pt $\sim$}}\raise 1pt\hbox {$>$}}}

\title[Abundance stratification in SN Ia - III. SN\,2003du]
{Abundance stratification in Type Ia supernovae - III. \\
The normal SN\,2003du}

\author[Tanaka et al.]{
\parbox[t]{\textwidth}{
Masaomi Tanaka$^{1}$\thanks{E-mail: masaomi.tanaka@ipmu.jp}, 
Paolo A. Mazzali$^{2,3,4}$, 
Vallery Stanishev$^{5}$, 
Immanuel Maurer$^{2}$,
Wolfgang E. Kerzendorf$^{6}$,
and
Ken'ichi Nomoto$^{1}$
}
\vspace*{6pt}\\
$^1${\it Institute for the Physics and Mathematics of the Universe, University of Tokyo, Kashiwa, Chiba 277-8568, Japan}\\
$^2${\it Max-Planck Institut f\"ur Astrophysik, Karl-Schwarzschild-Str. 1, 85748 Garching, Germany}\\
$^3${\it Scuola Normale Superiore, Piazza dei Cavalieri 7, 56126 Pisa, Italy}\\
$^4${\it Istituto Naz. di Astrofisica-Oss.\ Astron., vicolo dell'Osservatorio, 5, 35122 Padova, Italy }\\
$^5${\it CENTRA - Centro Multidisciplinar de Astrof\'isica, Instituto Superior T\'ecnico, Av. Rovisco Pais 1, 1049-001 Lisbon, Portugal}\\
$^6${\it Research School of Astronomy and Astrophysics, Mount Stromlo Observatory, Cotter Road, Weston Creek, ACT 2611, Australia}\\
}
\date{Accepted ---. Received ---}

\volume{000}

\begin{document}
\maketitle

\label{firstpage}

\begin{abstract}
The element abundance distributions in the ejecta of Type Ia supernova (SN) is
studied by modelling a time series of optical spectra of SN 2003du until $\sim
1$ year after the explosion. Since  SN 2003du is a very normal Type Ia SN both
photometrically and  spectroscopically, the abundance distribution derived for
it can be considered  as representative of normal Type Ia SNe.  We find that the
innermost layers are dominated by stable Fe-group elements, 
with a total mass of $\sim 0.2 \Msun$, 
which are synthesized through  electron capture.  Above the core of
stable elements there are thick \Nifs-rich layers. 
The total  mass of \Nifs\ is $0.65 \Msun$. 
The Si- and S-rich layers are located above the \Nifs-rich layers.
The dominant element in the outermost layers ($M_r > 1.1\Msun$,
$v > 13000$ \kms)  is O, with a small amount of Si. Little unburned
C remains, with an upper limit of 0.016 $\Msun$.
The element distributions in the ejecta are moderately mixed, 
but not fully mixed as seen in three-dimensional deflagration models.
\end{abstract}



\section{Introduction}

Type Ia supernovae (SNe Ia) are thought to be a thermonuclear explosion of
a C+O white dwarf (WD) in a close binary system
\citep[see][]{nomoto94Ia,hillebrandt00}. Since the ignition is triggered  when
the WD reaches a mass close to the Chandrasekhar limit, the properties of SNe Ia
are expected to be homogeneous.  SNe Ia are believe to synthesize a large amount
of \Nifs\ and other Fe-group elements during the explosion, and thus they are
one of the primary supply sources of Fe-group elements in the Universe. In
addition, the radioactive decay of \Nifs\ results in an extreme luminosity of
SNe Ia. This large luminosity, and their relative homogeneity  makes it possible
to use them as a standardisable candles with which distances can be measured in
the distant Universe \citep[\eg][]{phillips93, riess96, riess98, perlmutter99}.

However, the explosion mechanism of SNe Ia is still unclear. Although it is
widely accepted that the explosion starts as a deflagration, \ie a subsonic
burning flame \citep{nomoto76}, the subsequent evolution of the flame is under
debate: the flame might remain subsonic  (deflagration model, \eg
\citealt{nomoto84}), or experiences a transition to the supersonic regime
(delayed detonation model, \eg \citealt{khokhlov91}). These scenarios have been
studied in detail with multi-dimensional simulations
\citep[\eg][]{reinecke02,plewa04,roepke05,gamezo05,roepke07,jordan08,kasen09Ia}.

Different explosion models predict different element abundance distributions and
nucleosynthesis yields. For example, deflagration models tend to leave more
unburned elements than delayed detonation models, where the WD is burned almost
entirely. Three-dimensional deflagration models predict a mixed abundance
distribution, which is not seen in one-dimensional deflagration models
\citep{nomoto84} or in one- and multi-dimensional delayed detonation models
\citep{khokhlov91,gamezo05}. Thus, in order to place constrains on the explosion
mechanism, it is important that the abundance distribution and the 
nucleosynthesis yields are derived from observed SNe.

Modelling a time series of optical spectra of extragalactic SNe is a powerful
method to extract the abundance distribution in SNe Ia. At early epochs
(photospheric phase, $\texp \lsim 40$ days, hereafter $\texp$ denotes the time
since the explosion), the optical spectra of SNe Ia exhibit a pseudo-continuum
emission and P Cygni profiles superposed onto it. Since the absorption features
are formed by line scattering in the outer, optically thin layers, they carry
information on the abundances in the outer layers. In addition, since the SN
ejecta are expanding, the line-formation region moves inwards with time in the
mass and velocity coordinate. Following the time evolution of the spectra, the
abundances in different layers can thus be derived.

At late enough epochs (nebular phase, $\texp \gsim 150-200$ days), 
a SN shows an
emission-line spectrum. Most lines are forbidden lines of heavy elements. Since
these lines are formed in the innermost layers of the ejecta, we can derive the
abundances in the inner layers, which cannot be studied with the
photospheric-phase spectra.

Using the optical spectra, \citet[Paper I]{stehle05} derived the abundance
stratification in SN 2002bo. SN 2002bo is a normally luminous Type Ia SN (with a
B-band light curve (LC) decline rate $\Delta m_{15} = 1.13$ mag), 
but has higher line
velocities, \ie larger Doppler shifts at the absorption minima, than canonical
SNe Ia \citep{benetti04}. Paper I found that a large abundance of \Nifs\ and
intermediate mass elements (IMEs) such as Si and S is required in the outer
layers, and pointed out that mixed-out \Nifs\ can explain the fast rise of the
LC. \citet[Paper II]{mazzali0804eo} studied SN 2004eo. SN 2004eo has a
transitional luminosity between normal and subluminous SNe Ia and has a
relatively rapid decline rate of the light curve 
\citep[$\Delta m_{15} = 1.46$ mag,][]{pastorello07}.  A relatively small mass of 
Fe-group elements was obtained for SN\,2004eo in Paper II, suggesting that this 
is responsible for the rapid decline of the LC.

In this paper, we study the abundance distribution in SN 2003du
\citep{gerardy04, anupama0503du, stanishev07, marion09NIR}. SN 2003du is a very
normal Type Ia SN both photometrically and spectroscopically, and it can
therefore be considered as a representative of normal SNe Ia. SN 2003du is
classified as low velocity gradient (LVG) according to the scheme of
\citet{benetti05}, or as core-normal according to the scheme of \citet{branch06,
branch07}. This SN is an extremely well-studied object at optical and near
infrared (NIR) wavelengths, with an extremely good time sampling,  and one of
the rare cases in which  NIR spectra at late epochs have been obtained
\citep{hoeflich04, motohara06}.

This paper is organized as follows. In Section \ref{sec:method}, the methods of
analysis are described. The spectra at the photospheric and nebular phases are
modelled in Sections \ref{sec:early} and \ref{sec:late}, respectively. In
Section \ref{sec:LC}, the derived abundance distribution is tested against the
bolometric LC. The derived abundance distribution and the integrated yields are
shown in Section \ref{sec:abun}. In Section \ref{sec:con} we give conclusions.

Throughout the paper, the distance to SN 2003du is assumed to be $\mu=32.79$ mag
\citep{stanishev07}. The reddening in our Galaxy and in the host galaxy are
$E(B-V)=0.01$ mag and $0$ mag, respectively \citep{stanishev07}. We adopt a
bolometric rise time of 19 days, which was derived from the LC synthesis
\citep{stanishev07} and is consistent with the $B$-band rise time of 20.9 days.

The mass and mass fraction of radioactive \Nifs\ [$M$(\Nifs) and $X$(\Nifs),
respectively] are expressed by the value at the explosion ($\texp=0$). $M$(Fe)
[$X$(Fe)] and $M$(Ni) [$X$(Ni)] only represent the mass [mass fraction] of
stable Fe and Ni, respectively.

\section{Methods of Analysis}
\label{sec:method}

\subsection{Photospheric Phase}

The photospheric-phase spectra are studied using the Monte Carlo spectrum
synthesis code developed by \citet{mazzalilucy93}. Assumptions made in the code
are described in Appendix \ref{app:early}. For more details, see
\citet{mazzalilucy93,lucy99,mazzali00}. The code has been used for the modelling of various types
of SNe \citep[\eg][]{mazzali92, mazzali93, mazzali0097ef}.

The code calculates a synthetic spectrum for given inputs:
(1) a radial density profile, (2) a position of the photosphere in velocity 
(\ie photospheric velocity \vph)
\footnote{Since the SN ejecta are in homologous expansion 
($v = r/t$) at the time of observations,
the velocity can be used as a radial coordinate.
As the velocity is unchanged with time, 
it is more convenient than the radius itself.},
(3) an emergent luminosity $L$ and 
(4) element abundances above the photosphere.
For the density structure, we use the deflagration model W7 \citep{nomoto84}.
The validity of this choice is discussed in Section \ref{sec:abun}.
The other parameters are optimized so as to give the best fit of the 
observed spectra.

The luminosity $L$ is most strongly constrained by the observed absolute flux.
The photospheric velocity \vph\ is estimated primarily by using  the
overall colour of the observed spectra via the relation $L = 4 \pi v_{\rm ph}^2
\texp^2 \sigma T_{\rm eff}^4$,  where $T_{\rm eff}$ and $\sigma$ are the
effective temperature and the Stefan-Boltzmann constant, respectively. The
estimated velocity is further checked based on the reproduction of the observed
line velocities  (\ie Doppler shift at the absorption minimum of P Cygni
profiles,  see Fig. \ref{fig:vel}).

With the estimated $L$ and \vph, the abundance ratios above the photosphere are
optimized to reproduce the observed spectra. We start modelling from earlier
spectra for abundance stratification because the photosphere recedes with time
in the mass and velocity coordinates. Suppose that the photosphere recedes from
$v_1$ to $v_2$ ($v_1 > v_2$) in the time interval from $t_1$ to $t_2$ ($t_1 <
t_2$). First, the abundances at $v>v_1$ are optimized by fitting the observed
spectrum at $\texp=t_1$. At the later phase ($\texp=t_2$, $v_{\rm ph}=v_2$), 
the inner part begins to make a contribution to the spectrum. However, even at
$\texp=t_2$, the outer layers at $v>v_1$ still contribute to the spectrum. Thus,
we keep the abundance ratios at $v>v_1$ and optimize the abundances at $v_2 < v
< v_1$ by fitting the observed spectrum at $\texp=t_2$. Fitting the time series
of spectra, we get an abundance distribution as a function of velocity.

\subsection{Nebular Phase}

A nebular-phase spectrum is modelled under different assumptions from those made
for the photospheric-phase spectra. The assumptions made in our analysis are
described in Appendix \ref{app:late}. For more details, see \eg
\citet{axelrod80} and \citet{mazzali01}. This code has been used for the
modelling of supernovae \citep[\eg][]{mazzali01}.

The inputs are only the radial density profile and the element abundance
distribution. For the density structure, we use the W7 model as in the
photospheric phase. For the abundance distribution, the results of the
photospheric-phase modelling are used in the outermost layers ($v \gsim 15000$
\kms). At  intermediate layers ($8200$\kms\ $\lsim v \lsim 15000$ \kms), the
abundance distribution is constrained both from photospheric-phase spectra and
nebular-phase spectra. The abundances in the inner layers ($v < 8200 \kms$) are
constrained only from the nebular-phase spectra. The modelling is iterated until
a consistent distribution is obtained.

The total mass and distribution of \Nifs\ are especially important for the
modelling of nebular spectra because they determine the emergent luminosity and
the energy deposition in each layer. The total energy emitted by forbidden lines
of Fe is sensitive to the mass of \Nifs\ as well as to the total amount of Fe 
(\ie the sum of stable Fe and the decay product of \Nifs), which affects in
particular the ionization of Fe. Thus, although \Nifs\ is mostly Fe at the
nebular epoch, stable Fe and \Nifs\ products can be distinguished, albeit only
indirectly.

\begin{table}
\begin{center}
\caption{Parameters of the synthetic spectra at photospheric phases
and the calculated effective temperature and blackbody temperature}
\label{tab:param}
\begin{tabular}{rrrrrr}
\hline\hline
\noalign{\vspace{2pt}}
Epoch$^*$ & $\texp$  & $\log L$   & $v_{\rm ph}$ & $T_{\rm eff}$   &$T_{\rm B}$ \\
(day)   &  (day)  &  (erg s$^{-1}$)  &    (\kms)  & (K)    & (K)   \\
\hline
-12.8   &  8.1  & 42.43  &  11500& 8900 & 12700  \\
-10.9   & 10.0  & 42.69  &  10500& 9500 & 13500  \\
-7.8    & 13.1  & 42.97  &   9800& 10100 & 13800  \\
-5.8    & 15.1  & 43.08  &   9400& 10400 & 13800  \\ [2.5pt]

-4.0    & 16.9  & 43.11  &   9050& 10300 & 13400  \\
-1.9    & 19.0  & 43.17  &   8900& 10000 & 12500  \\
 0      & 20.9  & 43.18  &   8500& 9800  & 12000  \\
+1.2    & 22.1  & 43.18  &   8200& 9700  & 11800  \\ [2.5pt]

+4.3    & 25.2  & 43.12  &   7700& 9000 & 10700  \\
+7.2    & 28.1  & 43.08  &   7200& 8600 & 10000  \\
+10.0   & 30.9  & 43.00  &   6600& 8200 &  9400  \\
+17.2   & 38.1  & 42.78  &   4000& 8400 &  9800  \\
\noalign{\vspace{2pt}}
\hline
\end{tabular}\\
\end{center}
$^*$ Days since $B$ maximum
\end{table}

\section{Photospheric-Phase Spectra}
\label{sec:early}

Twelve photospheric-phase spectra of SN 2003du presented by \citet{stanishev07}
are modelled. The input parameters ($L$ and \vph) are summarized in Table
\ref{tab:param}. The effective temperature $T_{\rm eff}$ is calculated 
by the equation $L = 4 \pi v_{\rm ph}^2 \texp^2 \sigma T_{\rm eff}^4$.
The blackbody temperature $T_{\rm B}$ calculated with the code is
also shown in Table \ref{tab:param}.
The blackbody temperature is defined by 
$L_{\rm B} = 4 \pi  v_{\rm ph}^2 \texp^2 \sigma T_{\rm B}^4$, 
where the luminosity $L_{\rm B}$ in this equation 
is not the emergent luminosity but the luminosity 
emitted from the photosphere, including photons scattered backward into
the photosphere \citep{mazzalilucy93}.
Thus, $T_{\rm B}$ is always higher than 
the effective temperature ($T_{\rm eff}$).

\begin{figure}
  \includegraphics[scale=0.65]{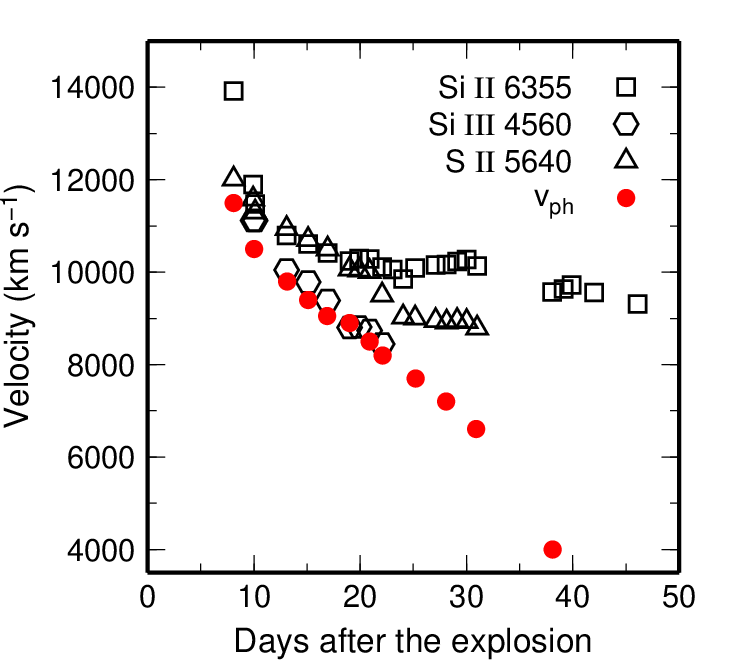} 
  \caption{
Line velocities measured in the observed spectra (open symbols,
\citealt{stanishev07}) compared with the photospheric velocities derived from
the spectral modelling (filled red circles). The photospheric velocity follows
the Si~{\sc III} line velocity until $\sim 20$ days after the explosion (around
maximum).
}
\label{fig:vel}
\end{figure}

In Fig. \ref{fig:vel}, the position of the photosphere is shown by red circles.
The observed line velocities of \SiII\ $\lambda$6355, \SiIII\ $\lambda$4560, and
\SII\ $\lambda$5640 are also shown for comparison \citep{stanishev07}. The time
evolution of the estimated photospheric velocity is  similar to that of the
\SiIII\ line velocities, which is the slowest among the three lines. 
This indicates that the slowest line velocity is the best observational indicator 
of the photospheric velocity.
Note that the estimated photosphere after maximum ($\texp \gsim 21$ days) 
is not very certain because the assumptions of the code become less 
reliable at later epochs (see Appendix \ref{app:early}).

In the following sections, the results of modelling at pre-maximum epochs (from
$-12.8$ to $-5.8$ days since $B$ maximum), maximum epochs (from $-4.0$ to $+1.2$
days) and post-maximum epochs (from $+4.3$ to $+17.2$ days) are presented.

\subsection{Pre-Maximum Epochs}

\begin{figure*}
  \includegraphics[scale=0.7]{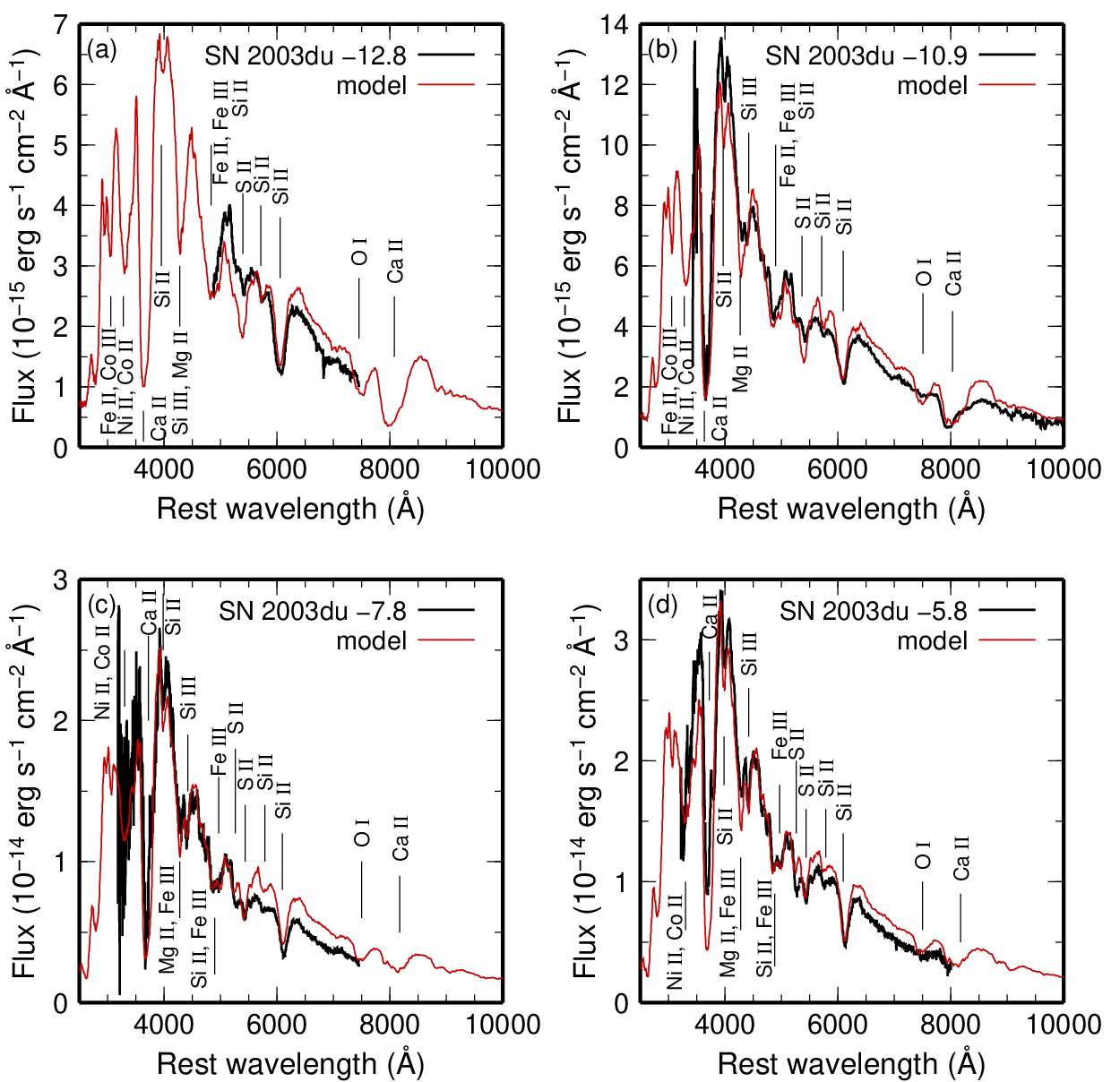}
  \caption{
The observed spectra of SN 2003du at the pre-maximum epochs (black) compared
with the synthetic spectra (red). Panels (a) - (d) show the spectra at $-12.8$,
$-10.9$, $-7.8$ and $-5.8$ days from $B$ maximum, respectively. Line
identifications shown in the figures are made only for the major contributions.
The synthetic spectra are reddened with $E(B-V)=0.01$ mag.
}
\label{fig:spec1}
\end{figure*}

\begin{figure*}
  \includegraphics[scale=0.7]{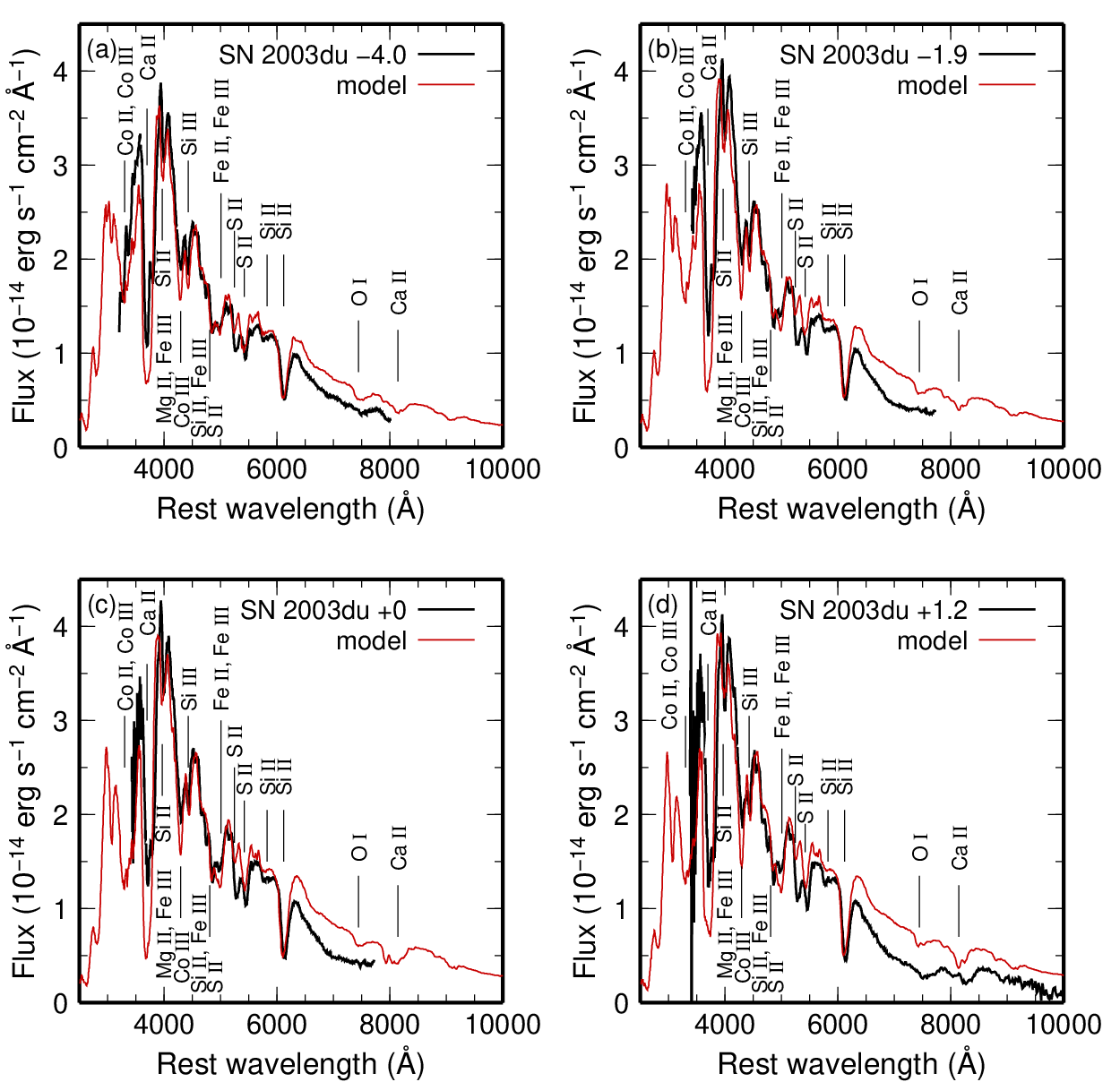}
  \caption{
Same as Fig. \ref{fig:spec1} but at the maximum epochs.
Panels (a) - (d) show the spectra at $-4.0$, $-1.9$, $0$ and $+1.2$ days
from $B$ maximum, respectively. 
}
\label{fig:spec2}
\end{figure*}

In Fig. \ref{fig:spec1}, the observed spectra at epochs from $-12.8$ to $-5.8$
days since $B$ maximum (black) are compared to the synthetic spectra (red).
The observed spectra at these epochs show a relatively well defined
pseudo-continuum and P Cygni profiles of IMEs and Fe-group elements. At these
epochs, the photospheric velocity decreases from 11500 to 9400 \kms.

{\it Fe-group elements}:  The presence of the Fe, Co and Ni lines clearly
indicates that some amount of these Fe-group elements are located in the outer
layers. Especially, since only a small amount of \Nifs\ has decayed into Fe at
such early epochs, the presence of Fe lines requires stable Fe in the
outer layers \citep{tanaka08Ia}. The mass fraction of stable Fe is found to be
$X$(Fe) $\sim 0.003$ - $0.005$.

{\it Calcium}:  In the spectrum at $-10.9$ days from $B$ maximum, the \CaII\
lines near 4000 \AA\ (\CaII\ H\&K) and 8000 \AA\ (\CaII\ IR triplet) show an
extremely high velocity ($v \sim 25000$ \kms). The origin of this line has been
discussed, including the interaction with the circumstellar matter (CSM) or high
velocity blobs \citep[\eg][]{hatano99, thomas04, gerardy04, mazzali0599ee,
mazzali05Ia, quimby06, tanaka06, altavilla07,garavini07}. 
In this paper, we do not consider the density enhancement with 
respect to the W7 model.
Keeping the W7 density structure, a mass fraction as high as $X$(Ca) $> 0.3$ 
is required at $v > 22000$\kms.

The density at the high velocity layers at $v>22000$ \kms\ in delayed
detonation models is higher than in the W7 model by a factor of $\sim 10$ (see
\eg Fig. 11 of \citealt{baron06}).  \citet{tanaka08Ia} studied the influence of
the density structure at the  outermost layers.  They found that the required
mass fraction of Ca is decreased by a factor of $\sim 10$ in the delayed
detonation models, but it is still higher than solar abundance (see their Table
3). The density enhancement may also possibly be caused by interaction with the
CSM \citep{gerardy04, mazzali0599ee}.

{\it Magnesium, Silicon, and Sulphur}:  Since the strongest Mg feature at 4200
\AA\ is contaminated by Fe lines, the mass fraction of Mg is not as well
constrained as those of Si, S, and Ca. If we assume that the Fe abundance is
constrained from the features around 4700 \AA, and we attribute any differences
to Mg, the mass fraction of Mg is $X$(Mg) $\sim 0.01 - 0.1$ at $v > 9400$ \kms.

SN 2003du has relatively narrow \SiII\ lines. To reproduce the profile of the
\SiII\ $\lambda$6355 line, the mass fraction of Si is required to be $X$(Si)
$\sim 0.3$. The dominant element in the layers at $v > 13000$ \kms\ is not Si, 
but more likely O. The distribution of Si is discussed in Appendix \ref{app:Si}.

The mass fraction of S is $X$(S) $= 0.1-0.2$ around $v = 10000$ \kms. Although
the synthetic spectra for the two earliest epochs show stronger lines than the
observations, the later spectra are nicely reproduced.

{\it Carbon and Oxygen}:  There is no clear C line in the observed spectra
except for the possible \CII\ $\lambda$6578 line at the emission peak of the
strongest \SiII\ $\lambda$6355 line \citep{stanishev07}. We derive an upper
limit for the C mass as $M$(C) $<$ 0.016 $\Msun$ at $v > 10500$ \kms\ 
(details are presented in Appendix \ref{app:C}). 
The small mass fraction and mass of C are consistent with the
results of previous works \citep{marion06, thomas07, tanaka08Ia,marion09NIR}.

The synthetic spectra show a clear \OI\ $\lambda$7774 line around 7500 \AA\
while the presence of such a strong O line is not that clear in the observed
spectra. Although the abundance of O is not well constrained, considering the
lower abundance of C, Mg, Si and S (see above), we conclude that the dominant
element in the outer layers is O. The mass fraction of O at $v>9400$ \kms\ is
$X$(O) $=0.05-0.85$, increasing towards the outer layers.

\subsection{Maximum Epochs}

\begin{figure*}
  \includegraphics[scale=0.7]{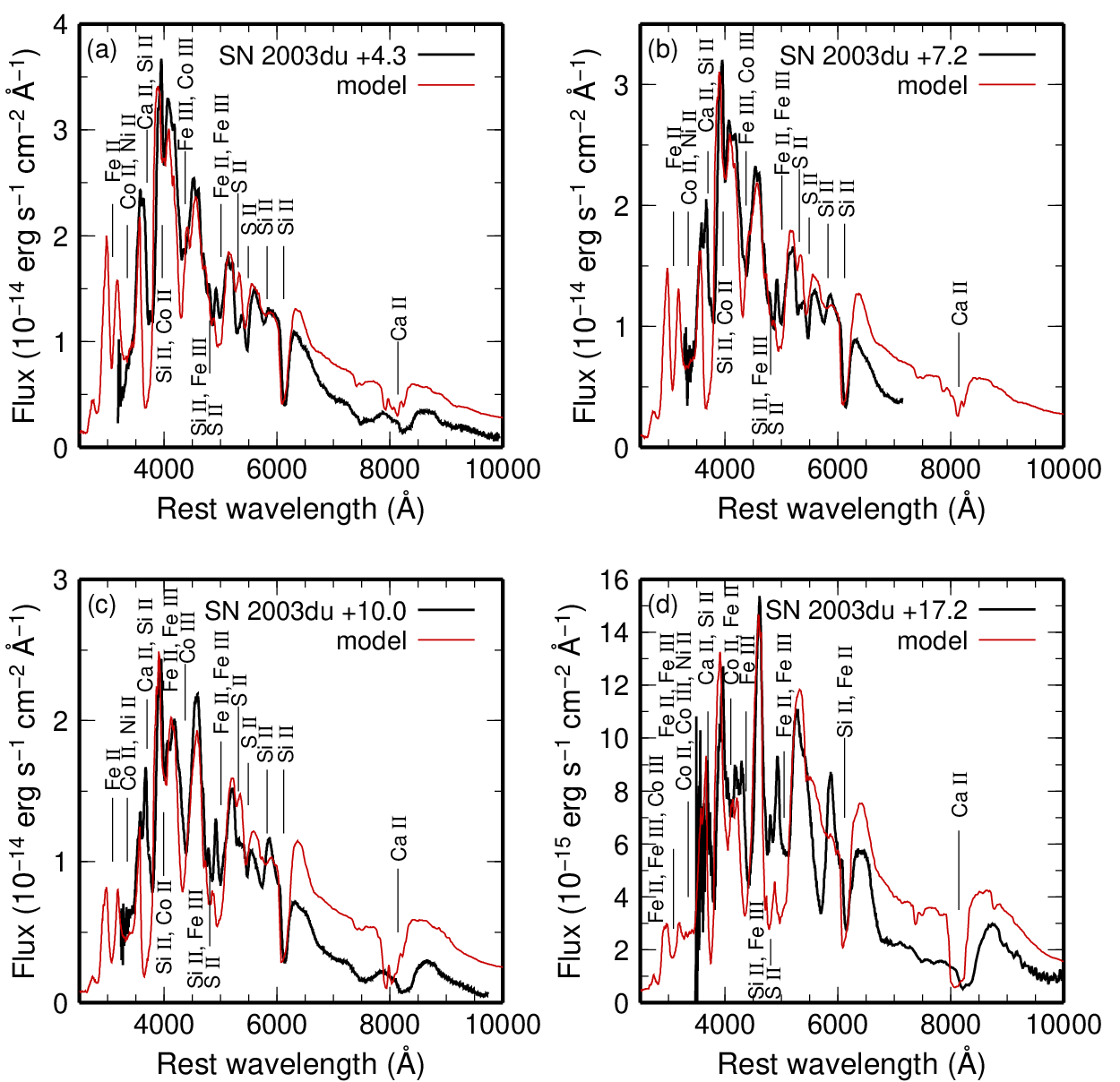}
  \caption{
Same as Fig. \ref{fig:spec1} but at the post-maximum epochs.
Panels (a) - (d) show the spectra at $+4.3$, $+7.2$, $+10.0$ and $+17.2$ days
from $B$ maximum, respectively.
}
\label{fig:spec3}
\end{figure*}

In Fig. \ref{fig:spec2}, the observed spectra at the epochs from $-4$ to $+1.2$
days since $B$ maximum (black) are compared to the synthetic spectra (red).
The element features present in the observed spectra are similar to those in the
pre-maximum spectra. At these epochs, the photospheric velocity decreases from
9050 to 8200 \kms. The flux of the synthetic spectra at the line-free region at
$\lambda > 6500$ \AA\ tends to be overestimated because of our assumption of
blackbody emission (see Appendix \ref{app:early}).

{\it Fe-group elements}: Around maximum, lines of Fe-group elements are present,
as in the pre-maximum epoch. A larger amount of \Nifs\ is required compared with
the pre-maximum epochs to block the emission in the near-UV more effectively.
The mass fraction of \Nifs\ is $X$(\Nifs) $= 0.5 - 0.7$ at $v = 8200 - 9050$
\kms.

At maximum epochs, the presence of Fe lines does not directly indicate the
presence of stable Fe because of the high abundance of the decay product of
\Nifs. At $B$ maximum ($\texp=20.9$ days), about 10 \%\ of \Nifs\ has already 
decayed into Fe, and thus an Fe abundance of $\sim 0.05 - 0.07$ is the product
of the decay of \Nifs. This is much larger than the abundance of stable Fe
required for the pre-maximum spectra, and thus we cannot distinguish the
contribution of stable Fe.

{\it Silicon and Sulphur}: The \SiII\ $\lambda$6355 line becomes narrower. The
Si distribution at the outer layers derived from the line profile in the 
pre-maximum spectra (Appendix \ref{app:Si}) nicely reproduce the maximum
spectra. In the spectrum at maximum, the line ratio of two \SiII\ lines 
($\lambda$5972 and 6355, $\mathcal{R}$(\SiII) in \citealt{nugent95}) is also
nicely reproduced, suggesting the estimates of the temperature and of the
ionization are reasonable (Table \ref{tab:param}, \citealt{hachinger08}).

The \SII\ lines around 5500 \AA\ are also reproduced nicely. Although the
photospheric velocity at this epoch is \vph $=8200 - 9050$ \kms, the \SII\
lines, as well as  the \SiII\ lines, are mainly formed at $v \sim 10000$ \kms\ 
(see Fig. \ref{fig:vel}). The ``detachment'' of the lines is caused by the
ionization effect, \eg \SIII\ is dominant near the photosphere while the
fraction of \SII\ increases towards higher velocities (see Fig. 9 of
\citealt{tanaka08Ia}).

\subsection{Post-Maximum Epochs}

In Fig. \ref{fig:spec3}, the observed spectra at epochs from $+4.3$ to $+17.2$
days since $B$ maximum are compared to the synthetic spectra. At these epochs,
the contribution from Fe-group elements is larger than at earlier epochs. The
assumption of blackbody emission becomes unreliable, and results in an
increasing overestimation of the flux at $\lambda > 6500$ \AA. Although the
photospheric velocity decreases from 7700 to 4000 \kms\ in our models (Table
\ref{tab:param}), these values are therefore not strongly constrained.

Since the assumption that there is no energy deposition above the photosphere is
also unlikely to hold at this epoch, we use the abundance distribution derived
from the modelling of the nebular spectrum (Section \ref{sec:late}), rather than
optimize the abundances using the spectra at post-maximum epochs.  Even with
this treatment, the synthetic spectra give reasonable agreement with the
observed spectra.

{\it Fe-group elements}: The \Nifs\ distribution derived from the nebular
spectrum reproduces the Ni, Co and Fe lines present in the observed spectra. 
The \NiII, \CoII\ and \FeII\ lines are responsible for the strong blocking of
the UV flux. The layers below 8200 \kms\ down to 4000 \kms\ (the photospheric
velocity  of the last spectrum, Fig \ref{fig:spec3}d) are rich in \Nifs, which
has a mass fraction of $X$(\Nifs)=$0.7-0.8$. Note that the Fe lines at 
$4500-5000$ \AA\ in the synthetic spectrum at $t=+17.2$ days are too strong. The discrepancy
is also seen in the model for SN 2004eo (Paper I), and it may result from our
assumption of no energy deposition in the atmosphere (see above).

{\it Sodium, Silicon, Sulphur, and Calcium}: In the last spectrum
(Fig. \ref{fig:spec3}d), there is a
strong absorption at 5600 \AA\ in the observed spectrum, which is expected to be
the \NaI\ D line. With our code, however, this line is not reproduced even if we
set $X$(Na) =0.1. This is already too high compared with the prediction by the
explosion models: the yield of Na is only $\sim 10^{-5} - 10^{-4} \Msun$
\citep[\eg][]{iwamoto99}.  This is the result of the overestimate of Na
ionization by our Monte Carlo code, which is a problem also for other NLTE codes
\citep[][and references therein]{mazzali97}, and it may be caused by
inaccuracies in the atomic data. Thus, we do not discuss total mass of Na in the
following section  (Section \ref{sec:abun}).

The strong \SiII\ and \SII\ lines are reproduced well in the spectra at $+4.3$
and $+7.2$ days from $B$ maximum (Fig. \ref{fig:spec3}a and \ref{fig:spec3}b).
These lines are formed in layers detached from the photosphere (see Fig.
\ref{fig:vel}) because of the high mass fractions of Si and S, and the
large fraction of \SiII\ and \SII\ in layers above the photosphere.

The \CaII\ H\&K and IR triplet lines in the synthetic spectra show too strong
high velocity component. They are caused by the high abundance at $v > 15000$
\kms\ derived from the earliest spectra. We believe that the constraints from
the earlier spectra are more reliable because the validity of the assumptions 
in the atmosphere is more certain at those epochs.

\section{Nebular-Phase Spectra}
\label{sec:late}

\begin{figure}
  \includegraphics[scale=0.70]{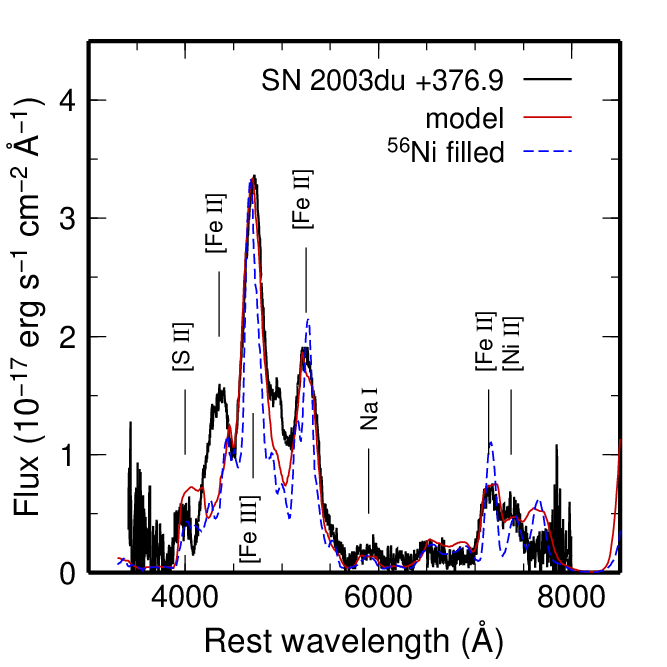} 
  \caption{The nebular spectrum of SN 2003du 
($+376.9$ days from $B$ maximum,  black line, \citealt{stanishev07})
compared with the synthetic spectra (color lines).
The red solid line shows the best fit model while the blue dashed
line shows the model without central region devoid of \Nifs.
}
\label{fig:late}
\end{figure}

A spectrum at $+376.9$ days since $B$ maximum \citep{stanishev07} is modelled to
derive the abundances in the deepest layers. A \Nifs\ distribution with a
decreasing mass fraction toward higher velocities ($v \gsim 10000$ \kms) nicely
fits the profile of the two strongest Fe lines (red line in Figure
\ref{fig:late}).  This is consistent with the constraints from the spectra at
the pre-maximum and maximum epochs.

The innermost layers are dominated by stable Fe. In particular, stable Fe and Ni
dominate the abundance at velocities inside of about 3000\,\kms, accounting for
a total mass of $\sim 0.2 \Msun$. Most of this is actually stable Fe. Limits to
the stable Ni mass can be set by the weakness of the [\NiII] emission near
7400\,\AA. As for stable Fe, this is required in the innermost regions in order
to keep an ionization balance between \FeII\ and \FeIII\ which allows the shape
of the two strong, Fe-dominated emissions at $4500-5000$ \AA\ to be reproduced, 
as well as their ratio. The bluer emission, near 4600\,\AA, is in fact dominated
by \FeIII, while the redder one, near 5200\,\AA, is dominated by \FeII. 

If a central region devoid of \Nifs\ is absent (blue dashed line), both Fe
features look very sharp, unlike the observations. In order to reduce the flux,
the \Nifs\ distribution must be limited to a smaller region, and the overall
higher density of the inner region leads to a lower Fe ionization, so that the
ratio of the two Fe lines in the blue is no longer correct: the \FeII-dominated
line (near 5200 \AA) becomes too strong, as shown in the model in blue dashed
line in Figure \ref{fig:late}. On the other hand, the presence of stable Fe acts
as a coolant, keeping the ionization degree to a level which yields reasonable
line intensities (red line).

\section{Bolometric Light Curve}
\label{sec:LC}

The derived abundance distribution is tested against the bolometric LC
\citep{stanishev07}. SN 2003du is an extremely well-observed SN, and the
bolometric LC was constructed with a small uncertainty, including the
contribution of the NIR \citep{stanishev07}.

For the modelling, we use a Monte Carlo LC code based on that developed in 
\citet{cappellaro97}. 
The code follows the emission and propagation of
$\gamma$-rays and positrons from \Nifs\ and \Cofs\ decay. Both $\gamma$-rays and
positrons are treated with an effective opacity (\ie the positrons are not
assumed to deposit in situ). The opacities adopted are 0.027 cm$^2$ gm$^{-1}$
for the $\gamma$-rays and 7 cm$^2$ gm$^{-1}$ for the positrons
\citep{axelrod80}.
For optical opacity, we use an analytic formula introduced 
by \citet{mazzali01Ia,mazzali07Ia}, which includes the composition dependence
(Fe-group elements and other elements) and a time dependence
\citep{hoeflich96Ia}.

\begin{figure}
  \includegraphics[scale=0.7]{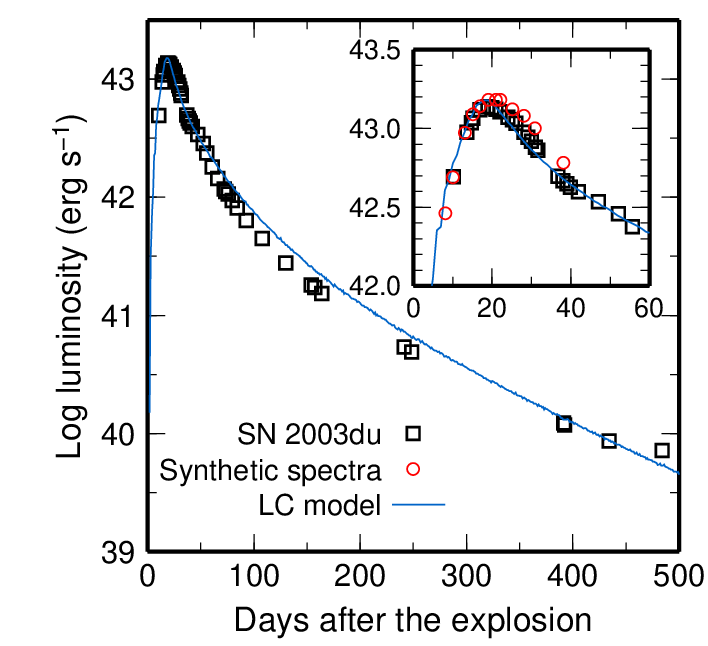} 
  \caption{Bolometric LC of SN 2003du (black squares, \citealt{stanishev07}) 
compared with the bolometric LC model (thin line) and 
the input bolometric luminosities in the spectral modelling (red circles).
  }
\label{fig:LC}
\end{figure}

Fig. \ref{fig:LC} shows the comparison between the observed bolometric LC (black
squares) and the computed LC (blue line). The input luminosities for the
photospheric-phase spectra are also  plotted (red circles). The model LC matches
well with the observed LC from before maximum until $> 1$ year. This suggests
that the derived abundance distribution, especially that of \Nifs, is
reasonable. The bolometric LC derived from the synthetic spectra at epochs after
maximum deviate progressively from the observational points, reflecting the
increased production of spurious flux redwards of 6500 \AA\ in the synthetic
spectra (see Sect. 4).

\section{Abundance Distribution and Integrated Yields}
\label{sec:abun}

\begin{figure*}
  \includegraphics[scale=0.65]{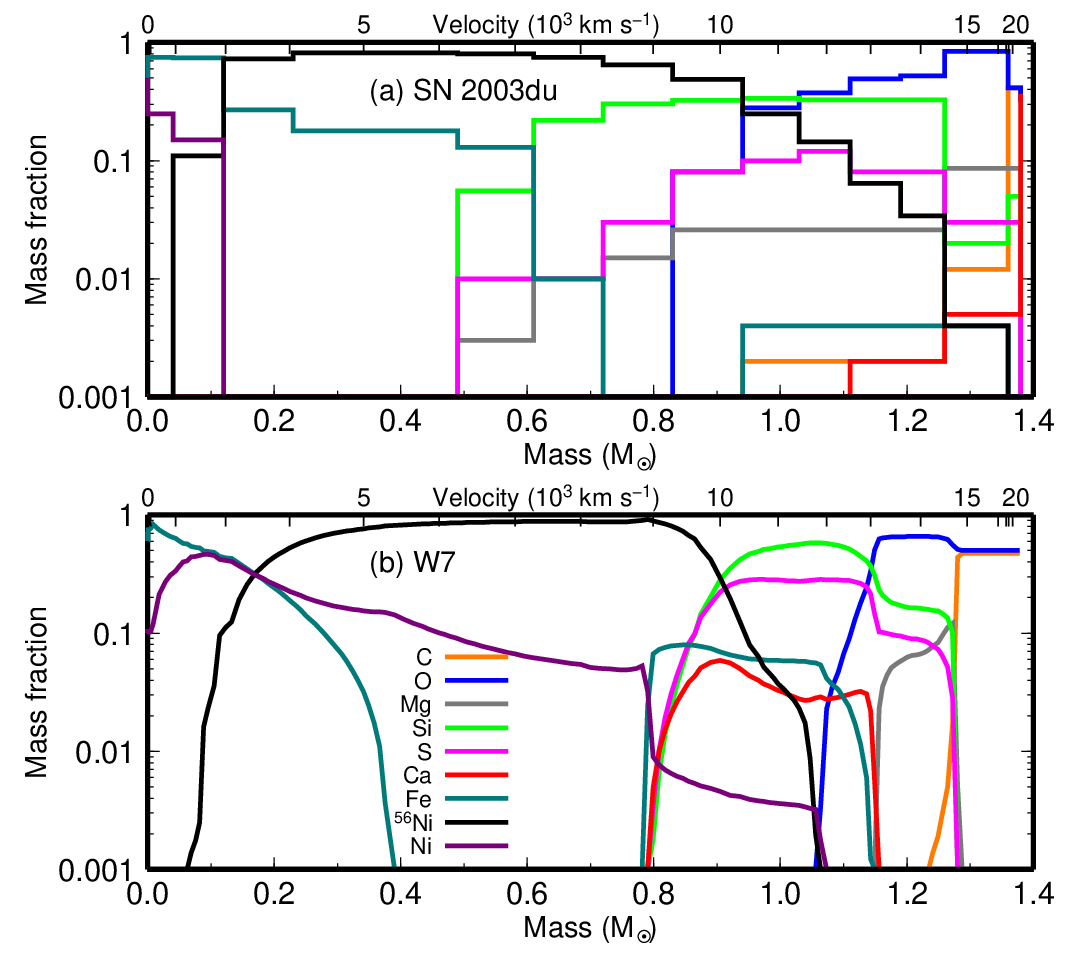} 
  \caption{Abundance distribution in the mass coordinate 
derived from the spectral modelling (a) compared with that of W7 (b).
``\Nifs'' represents the mass fraction of \Nifs\ at $\texp=0$ and ``Fe'' and 
``Ni'' only represent mass fractions of stable Fe and stable Ni, respectively.
  }
\label{fig:abunmass}
\end{figure*}

\begin{figure*}
  \includegraphics[scale=0.65]{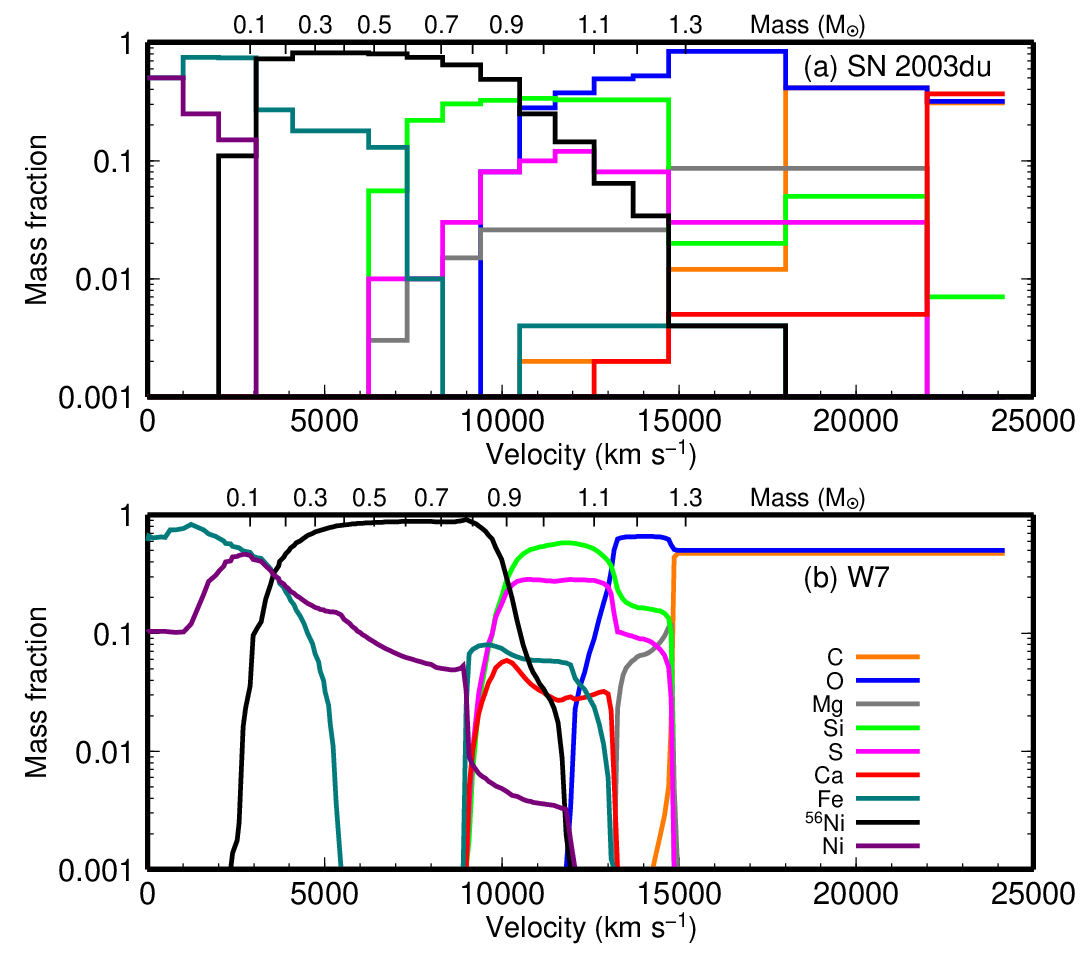} 
  \caption{
Same as Fig. \ref{fig:abunmass} but in the velocity coordinate.
}
\label{fig:abunvel}
\end{figure*}

\begin{figure*}
  \includegraphics[scale=0.70]{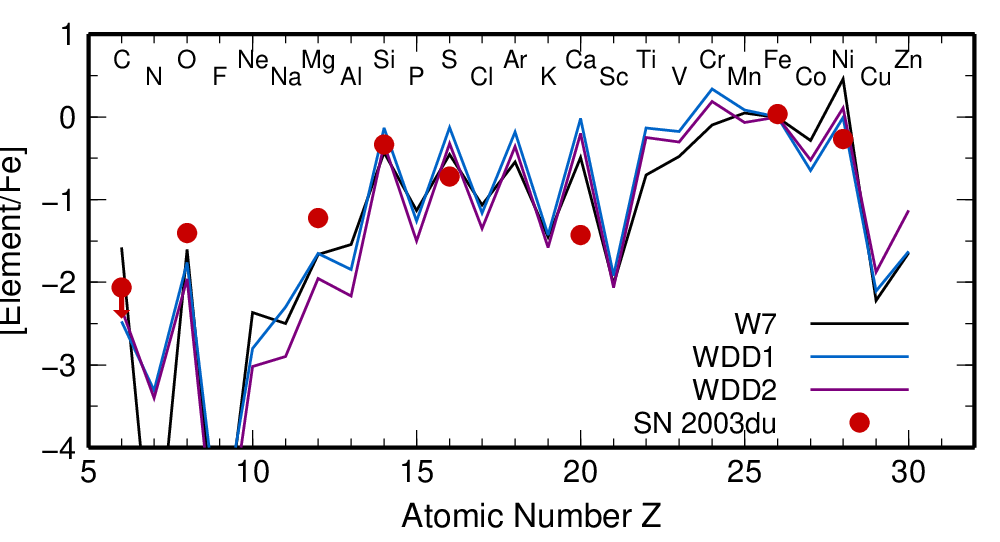} 
  \caption{ 
Abundance ratios, \ie
$[{\rm Element / Fe}] = {\rm log}_{10}[M/M({\rm Fe})] - {\rm log}_{10}[X/X({\rm Fe})]_{\odot}$
of SN 2003du (red circles), W7 (black line, \citealt{nomoto84}), 
WDD1 (blue line, \citealt{iwamoto99}), and WDD2 (purple line, \citealt{iwamoto99}).
Note that, in this figure, the yields after the radioactive decay are shown,
\ie Fe includes both of stable Fe and the decay product of \Nifs, 
and Ni includes only stable Ni.
}
\label{fig:yield}
\end{figure*}
\subsection{Abundance Distribution}
\label{sec:distribution}

The abundance distribution derived from the modelling of the photospheric- and
nebular-phase spectra is shown in Figs. \ref{fig:abunmass} and
\ref{fig:abunvel}. These two figures show the mass fraction of each element as a
function of mass and velocity, respectively. For comparison, the abundance
distribution of a deflagration model \citep[W7,][]{nomoto84} is also shown 
(Figs \ref{fig:abunmass}b and \ref{fig:abunvel}b).  Note the line with ``\Nifs''
represents the mass fraction of radioactive \Nifs\ at the time of the explosion
($\texp=0$), and that ``Fe'' and ``Ni'' denotes only stable elements.

At the innermost layers ($M_r < 0.1 \Msun$, $v < 3000$ \kms), stable Fe 
dominates (Section \ref{sec:late}). This is also suggested by the NIR spectrum
at late phase \citep{hoeflich04,motohara06}. The existence of stable elements at
the innermost layers is consistent with the explosion model. These elements are
synthesized via electron capture reactions at high densities, and thus are
typically located in the innermost layers.

Above the layer of stable elements, there is a \Nifs\ -dominated layer ($M_r =
0.1-0.9 \Msun$, $v=3000-10000$ \kms). This is fairly consistent with the
distribution of W7. However, in SN 2003du a moderate amount of \Nifs\ is
distributed also in the outer layers, which is important for the blocking of UV
light in the early spectra. This was also true for SN\,2002bo (Paper I). At $M_r
> 0.9 \Msun$ ($v>10000$ \kms), there is $\sim 0.01 \Msun$ of \Nifs. 

The layers dominated by Si and S are located above the \Nifs\ -rich layers.
Compared with W7, Si has a broader distribution in the mass/velocity
coordinates. The moderate amount of Si in the \Nifs\ -rich layers is required to
explain the Si absorption line in the spectra after maximum. Even at the outer
layers at $v>15000$ \kms\ ($M_r > 1.3 \Msun$), $X$(Si) $=0.01-0.1$ is also
required (Appendix \ref{app:Si}). This is in contrast to the W7 deflagration
model, where C+O remain unburned at $v\gsim 15000$ \kms.

Above the Si and S-rich layers, the dominant element is O ($M_r > 1.1 \Msun$,
$v>13000$ \kms). Although this is not directly constrained, a high abundance of
Si would not be consistent with the observed line profile (Appendix
\ref{app:Si}) and the mass fractions of C, Mg, S, and Ca are also not larger
than O.

The mass of unburned C is quite small, $M$(C) $< 0.016 \Msun$ at $v > 10500$
\kms (Appendix \ref{app:C}). The small C mass indicates that the progenitor C+O
WD was almost completely burned. In the outermost layers ($M_r > 1.3 \Msun$,
$v>15000$ \kms) only C-burning occurs, and elements heavier than O are not
synthesized effectively.
Note that there is no constraint on C above $v=18000$ \kms, 
and a mass fraction as high as $X({\rm C})=0.5$ can be allowed.
(see Appendix \ref{app:C}).

It is also possible that the C abundance in the outermost
layers of the progenitor WD is already suppressed in the pre-SN stage. In the
single-degenerate scenario, the WD grows by mass accretion from a companion
star. In the case of weak He flashes, their nucleosynthesis products can accrete
onto the WD. Since the products of the He flashes are not always C and O but
could also be heavier elements \citep[\eg][]{hashimoto83}, the outermost layers
of the WD do not necessarily consist  only of nearly equal mass fractions of C
and O.
The heavier elements synthesized in the pre-SN stage 
can also be responsible for the high velocity features of the Si and Ca lines
present in the earliest spectra \citep{tanaka08Ia}.

The overall abundance distribution is not consistent with the almost fully mixed
distribution seen in three-dimensional deflagration models
\citep{travaglio04,roepke05}. But the boundaries between stable Fe/\Nifs,
\Nifs/IME, IME/O-rich layers are less clear than in the one-dimensional models.
This requires moderate degree of mixing during the explosion \citep{woosley07}.

The sensitivity of the results on the parameters has been studied in Paper II.
In the photospheric phase, the synthetic spectra are sensitive to the velocity
at the inner boundary and the element mass fractions. Variations in the mass
fractions of up to 25 \% are possible, while the synthetic spectra are more
sensitive to variations of the velocity: a change of more than 15\% can
noticeably affect the fits (see Paper II). In the nebular phase, no given inner
boundary is required and the constraints on the mass fractions are less
uncertain.

Except for the elements shown in Figures \ref{fig:abunmass} and \ref{fig:abunvel},
there must be some amount of Ti and Cr in the ejecta.
For photospheric-phase spectra, these elements are responsible 
for the features or the flux level around $3000-3500$ \AA.
The mass fraction of the sum of Ti and Cr is about $0.001-0.01$ (Papers I and II).
However, since this wavelength range is not covered in most spectra 
used in this paper,
we refrain from discussing the distribution and total yields of these elements.

\subsection{Integrated Yields}
\label{sec:yield}

\begin{table*}
\caption{The total mass of the elements in SN~2003du derived from spectral modelling
and nucleosynthesis products of different SN~Ia explosion models} 
\label{tab:mass}
\begin{tabular}{lrrrrrrr}
\hline \hline
Species  & SN 2003du   & W7$^1$    & WDD1$^2$ & WDD2$^2$  & b30\_3d\_768$^3$ & Y12$^4$ & O-DDT$^5$\\
\hline		         		     		     
    C    &    $<$1.6E-02$^*$& 4.83E-02  & 5.42E-03 & 8.99E-03  & 2.78E-01  & 8.73E-03 & 3.49E-03 \\
    O    &    2.3E-01       & 1.43E-01  & 8.82E-02 & 6.58E-02  & 3.39E-01  & 1.07E-01 & 1.51E-01 \\
    Na   &    ---$^{**}$    & 6.32E-05  & 8.77E-05 & 2.61E-05  & 8.65E-04  & ---$^{***}$ & 5.87E-05 \\
    Mg   &    2.4E-02       & 8.58E-03  & 7.69E-03 & 4.52E-03  & 8.22E-03  & 8.70E-2 & 2.16E-02 \\
    Si   &    2.1E-01       & 1.57E-01  & 2.74E-01 & 2.07E-01  & 5.53E-02  & 1.27E-1 & 2.87E-01 \\
    S    &    4.8E-02       & 8.70E-02  & 1.63E-01 & 1.24E-01  & 2.74E-02  & 7.03E-2 & 1.27E-01\\
    Ca   &    1.4E-03       & 1.19E-02  & 3.10E-02 & 2.43E-02  & 3.61E-03  & 1.82E-2 & 1.70E-02 \\
    Ti   &     ---$^{**}$  & 3.43E-04  & 1.13E-03 & 1.02E-03  & 8.98E-05  & 1.41E-5 &  8.72E-04\\
    Cr   &     ---$^{**}$  & 8.48E-03  & 2.05E-02 & 1.70E-02  & 3.19E-03  & 2.96E-4 &  1.04E-02 \\
  Fe$^a$ &    1.8E-01       & 1.63E-01  & 1.08E-01 & 1.02E-01  & 1.13E-01  & 6.50E-3$^b$ & 1.11E-01\\
\Nifs    &    6.5E-01       & 5.86E-01  & 5.64E-01 & 6.90E-01  & 4.18E-01  & 9.26E-1$^c$ & 5.40E-01 \\
Ni$^d$   &    2.4E-02       & 1.26E-01  & 3.82E-02 & 5.87E-02  & 1.06E-01  & ---$^{***}$ & 8.05E-02 \\
\hline
\end{tabular}\\
$^{*}$Upper limit at $v>10500$ \kms. $^{**}$Not constrained. $^{***}$Not given.\\
$^a$Stable isotopes except for \Fefs\ from \Cofs\ decay. \\
$^b$Only $^{52}$Fe.\\
$^c$The sum of Fe-group elements except for $^{52}$Fe, and thus, 
the upper limit for \Nifs\ \citep[see][]{plewa07}.\\
$^d$Stable isotopes $^{58}$Ni, $^{60}$Ni, $^{61}$Ni, $^{62}$Ni, and $^{64}$Ni.\\
$^{1}\,$\citet{nomoto84}, $^{2}\,$\citet{iwamoto99},
$^{3}\,$\citet{travaglio04}, $^{4}\,$\citet{plewa07}, $^{5}\,$\citet{maeda10nuc}
\end{table*}

The ejected mass of each element in SN 2003du was calculated by integrating the
abundance distribution assuming the W7 density structure. Results are summarized
in Table \ref{tab:mass}. For comparison, the nucleosynthetic yields of the
following models are also shown: one-dimensional deflagration model
\citep[W7,][]{nomoto84}, one-dimensional delayed detonation models (WDD1 and
WDD2 \footnote{These two delayed detonation models have different transition 
densities $\rho = 1$ and $2 \times 10^7 \ {\rm g\ cm^{-3}}$ for WDD1 and  WDD2,
respectively.}, \citealt{iwamoto99}), three-dimensional deflagration model
\citep[b30\_3d\_768,][]{travaglio04}, two-dimensional detonating failed
deflagration model \citep[Y12,][]{plewa07}, and two-dimensional  off-center
delayed detonation model \citep[O-DDT,][]{maeda10nuc}.

In Figure \ref{fig:yield}, the derived yields are compared with 
those of the one-dimensional models 
in the abundance ratio to Fe, normalized by the solar abundances, \ie
$[{\rm Element / Fe}] = {\rm log}_{10}[M/M({\rm Fe})] - {\rm log}_{10}[X/X({\rm Fe})]_{\odot}$.
Here, the yields after the radioactive decays are shown, \ie
Fe means the sum of stable Fe and the decay product of \Nifs,
and Ni includes only stable Ni.

In SN 2003du, the ejected mass of neutron-rich elements is estimated to be 
$\sim 0.2 \Msun$. They are synthesized by electron capture reactions and 
mostly distributed in the innermost layers. The mass of \Nifs\ is $0.65 \Msun$,
which is the largest among all synthesized elements. The mass of IMEs (Mg, Si,
S, and Ca) is $\sim 0.28 \Msun$ in total. The mass of O is estimated to be
$0.23 \Msun$, and is predominantly located in the outer layers. There is no
sign of C, and the upper limit is $M$(C) $< 0.016 \Msun$ at $v > 10500$ \kms.

From these yields, the validity of the W7 density structure can be tested. We
use the formula for the kinetic energy of SNe Ia: $E_K = [1.56 M({\rm^{56}Ni}) +
1.74 M({\rm stable\ Fe}) + 1.24 M({\rm IME}) -0.46] \times 10^{51}$ erg
\citep{woosley07}, where $M$(\Nifs), $M$(stable Fe) and $M$(IME) are the ejected
mass of \Nifs, stable Fe-group elements and the IMEs, respectively. Substituting
the element masses derived for SN 2003du, we derive an ejecta kinetic energy of
$E_K = 1.25 \times 10^{51}$ erg, which is in a good agreement with the kinetic
energy of the W7 model ($E_K \sim 1.3 \times 10^{51}$). Thus, the element masses
derived from the modelling under the assumption of the W7 density structure are
self-consistent.

We first compare the integrated yields with those of one-dimensional models 
(Table \ref{tab:mass}).  The total mass of stable Fe-group elements estimated
for SN 2003du is reasonably consistent with that of the one-dimensional
explosion models. The amount of neutron-rich stable elements in the central
region is sensitive to the deflagration speed, which affects the efficiency of
the electron captures. 
Thus, the good agreement implies that the deflagration speed in the
explosion models is reasonable.

The mass of \Nifs\ in SN 2003du is also consistent with that of the
one-dimensional deflagration and delayed detonation models. In the delayed
detonation models, a larger amount of \Nifs\ suggests an earlier transition to a
detonation. Thus, if the delayed detonation model is the correct scenario, a
transition density close to that of the models WDD1 and WDD2 is reasonable for
SN 2003du.

The total mass of IMEs estimated for SN 2003du is smaller than in the delayed
detonation models, while the mass of O in SN 2003du is larger. This suggests
that O burning in SN 2003du was not as strong as in the delayed detonation
models. However, the unburned C mass in the outermost layers in SN 2003du is 
smaller than in the deflagration models. Thus, the outermost part of the C+O WD
is burned more intensely than in the deflagration model (see also Section
\ref{sec:distribution}).

Next, we compare the results with multi-dimensional models.
The mass of \Nifs\ in SN 2003du is clearly larger than in the three-dimensional
deflagration model \citep{travaglio04}. 
The three-dimensional deflagration model also predicts more 
unburned C, which is not consistent with the small C mass derived for SN 2003du.
Thus, also from the nucleosynthetic point of view, the three-dimensional
deflagration model is unlikely to be a reasonable model for a normal SNe Ia such
as SN 2003du.

It is interesting to compare the yields with the detonating failed deflagration
model \citep{plewa07} and the off-center delayed detonation model 
\citep{maeda10nuc},  because these models may explain the kinematic offset in
the central region suggested by the NIR spectra at late phases
\citep{hoeflich04,motohara06,maeda10neb}.  The detonating failed deflagration
model \citep{plewa07} predicts too much \Nifs\ compared with SN 2003du.  Thus,
this model does not seem a likely model for normal SNe Ia, as already cautioned
by \citet{meakin09}. The yields of the two-dimensional off-center delayed
detonation model  \citep{maeda10nuc} are in good agreement with those derived
for SN 2003du. It is, however, noted that the model predicts \Nifs\ synthesized
near the center, while the existence of a region free of \Nifs\ is  suggested
from our one-dimensional analysis of the nebular phase spectrum (Section
\ref{sec:late}).

\subsection{Comparison with SNe 2002bo and 2004eo}

In this section, we compare the abundance distributions and 
integrated yields derived for SN 2003du with those derived for 
SNe 2002bo (Paper I) and 2004eo (Paper II).
SN 2002bo is a normally luminous SN ($\Delta m_{15} = 1.13$ mag), 
but has high line velocities 
and a high velocity gradient at early phases \citep{benetti04}.
SN 2004eo has a transitional luminosity between normal and 
subluminous SNe Ia, with a relatively rapid decline of the light curve
\citep[$\Delta m_{15} = 1.46$ mag,][]{pastorello07}.

These three SNe have in common an innermost layer of stable Fe-group elements
which are synthesized by electron capture. The mass of this region $\sim 0.2
\Msun$ is similar in the  three SNe studied thus far \citep[see
also][]{mazzali07Ia}, although the mass of \Nifs\ is different, 0.65, 0.52, and
0.43 $\Msun$ for SNe 2003du, 2002bo, and 2004eo, respectively.

The most significant difference between SNe 2003du and 2002bo is in the
outermost layers.  In SN 2002bo Si reaches the outer layers (Paper I).  The
abundance of Si in the outer layers of SN 2002bo is larger than in SN 2003du. 
This is the primary reason why Si line velocities in SN 2002bo are higher than
in SN 2003du, as demonstrated in Appendix \ref{app:Si}.  \citet{tanaka08Ia}
also suggested that the difference between  high-velocity and low-velocity SNe
Ia is caused by the difference in the Si abundance.

Alternatively, the difference can also be caused by the difference in the
density structure in the outer layers. \citet{maeda10nat} pointed that the
difference in the velocity  (and/or velocity gradient) is caused by
viewing-angle effects in an aspherical explosion that has different density
structures depending on the viewing angle.

SN 2002bo also has a more extended \Nifs\ distribution than in SN 2003du. This
explains the rapid rise of the LC of SN 2002bo (Paper I). \citet{pignata08}
suggest that the rising time of the LC of  SNe with a high velocity gradient is
generally shorter than that  of SNe with a low velocity gradient. Thus, the
extended \Nifs\ distribution may be a common property of high-velocity SNe.

Interestingly, SNe 2003du and 2004eo have a similar distribution 
of Si, O, and C above $v=10000$ \kms\ ($M_r = 0.9 \Msun$).
This indicates that the different shapes of the LCs of these
two SNe result from the difference in the inner layers.
In fact, a difference between these SNe is seen in the mass 
and mass fraction of \Nifs.
A smaller amount of \Nifs\ is responsible for the fainter luminosity 
and more rapid decline of SN 2004eo (see also Paper II).

\section{Conclusions}
\label{sec:con}

We have studied the element abundance distribution in SN 2003du by modelling
the optical spectra at the photospheric and nebular phases. These are used to
place constraints on the abundances at the outer and inner layers, respectively.
Since SN 2003du is a normal Type Ia SN, both photometrically and
spectroscopically, the abundance distribution derived for SN 2003du can be
considered as representative of normal Type Ia SNe.

The abundance distribution and the integrated yields in SN 2003du can be
summarized as follows. The innermost layers are dominated by stable,
neutron-rich elements, with a total mass of $\sim 0.2 \Msun$. The yield is
roughly consistent with the results of one-dimensional models.  This may imply
that the deflagration speed assumed in the models is reasonable.

Above the stable elements, there is a thick \Nifs-rich layer. The total mass of
\Nifs\ is $0.65 \Msun$, which is consistent  with the one-dimensional
deflagration model and the one-dimensional delayed detonation models  with
transition density $\sim 1-2 \times 10^7 \ {\rm g\ cm^{-3}}$. The \Nifs\ mass is
larger than in the three-dimensional deflagration model and smaller than in the
detonating failed deflagration model. 

The layers dominated by Si and S are located above the \Nifs\ -rich layers.
The dominant element at $M_r > 1.1\Msun$ ($v > 13000$ \kms) is O. The
progenitor C+O WD is almost entirely burned, and the mass of leftover unburned C
is quite small ($< 0.016 \Msun$ at $v>10500$ \kms). The small mass of unburned C
is consistent with the delayed detonation model. However, the mass of O derived
for SN 2003du is larger than that in the delayed detonation model. These results
suggest that in the outermost layers C-burning is more intense than in the
deflagration model, but O burning is not as strong as in the delayed detonation
models.

The boundary between stable Fe/\Nifs,
\Nifs/IME, IME/O-rich layers is not as sharp as in the one-dimensional models,
suggesting a moderate degree of mixing. However, the strong mixing seen in the
multi-dimensional deflagration model is not preferable.

\vspace{3mm}
\noindent
This work has made use of the SUSPECT supernova spectral archive.
V.S. is financially supported by FCT Portugal under 
program Ci\^{e}ncia 2008.
This research has been supported in part by World Premier
International Research Center Initiative, MEXT, Japan.




\appendix

\section{Assumptions in the Code}

\subsection{Photospheric Phase}
\label{app:early}

The spectrum synthesis code for the photospheric phases assumes a sharply
defined spherical photosphere as an inner boundary. At the inner boundary,
blackbody radiation is assumed. Since the temperature structure [$T_R (r)$] is
unknown at first, the calculation starts with an assumed temperature. Using this
temperature, the ionization and excitation in the optically thin atmosphere
above the photosphere are calculated considering the dilution of the radiation.
The degree of dilution $W$ is defined by the equation  $J = WB(T_R)$ in our
code.

The excitation (the population of an excited level $n_j$; $j=1$ for the 
ground state) is computed as in \citet{abbott85} and \citet{lucy99}:
\begin{equation}
\frac{n_j}{n_1} = W \frac{g_j}{g_1} e^{-\epsilon_j /k_B T_R},
\label{eq:level}
\end{equation}
where $g_j$ and $\epsilon_j$ are the statistical weight  
and the excitation energy from the ground level, respectively.

For the ionization, the nebular approximation, which is similar to that made in
the studies of stellar winds \citep{lucy70, abbott85} or planetary nebulae
\citep{gurzadyan97}, are adopted with  the modifications for SNe
\citep{mazzalilucy93}:
\begin{equation}
\frac{N_{i+1}N_e}{N_i} 
= [\delta \zeta + (1-\zeta)W] W \left( \frac{T_e}{T_R} \right)^{1/2} 
 \left( \frac{N_{i+1}N_e}{N_i}  \right)^{*}_{T_R},
\label{eq:ion}
\end{equation}
where $N_e$ and $T_e$ are the electron density and the electron temperature,
respectively. In the equation, $\delta$ and $\zeta$ are the correction factors
for the optically thick continuum shorter than the \CaII\ ionization edge and
the fraction of recombinations that go directly to the ground state,
respectively \citep{mazzalilucy93}. The first and second terms in the first
square brackets on the right hand side represents the excitation from the ground
and excited levels, respectively. The last term on the right hand side is the
ionization computed from the Saha equation with $T_R$. For the electron
temperature, $T_e = 0.9 T_R$ is crudely assumed \citep{klein78,abbott85}.

With the ionization and excitation states, line optical depths are calculated
under the Sobolev approximation, which is a sound approximation in rapidly
expanding envelope with large velocity gradients \citep{castor70, lucy71}. A
number of photon packets are traced above the photosphere in spherical
coordinates, taking into account electron scattering and line scattering. For
line scattering, the effect of the line branching is included \citep{lucy99,
mazzali00, pinto00}. The stream of the photon packets gives the flux at each
radial mesh and a frequency moment is calculated by
\begin{equation}
\bar{\nu} = \frac{\int \nu J_{\nu} d\nu}{\int J_{\nu} d\nu}.
\end{equation}
The radiation temperature is estimated from this frequency moment via
\begin{equation}
\bar{x} = \frac{h\bar{\nu}}{k_B T_R}.
\end{equation}
Here $\bar{x}$ represents for the mean energy of the blackbody radiation
[$\bar{x} = (h/ k_B T_R)(\int \nu B_{\nu} d\nu/\int B_{\nu} d\nu) =  3.832$].

Using this new temperature structure, the ionization and excitation are updated.
Then, the next Monte Carlo ray tracing starts. This procedure is continued until
the temperature converges. Finally, the emergent spectrum is calculated using a 
formal integral \citep{lucy99}.

It is known that these assumptions give a good approximation to the results of
detailed NLTE calculations \citep{pauldrach96}. However, since no energy
deposition by \Nifs\ and \Cofs\ is assumed  above the photosphere, the
assumptions become unreliable as a large part of the \Nifs-rich layer becomes
optically thin. In addition, the assumption of a sharp photosphere becomes
progressively invalid after maximum brightness. Thus, the quantities derived by
fitting the spectra at later epochs (especially after maximum brightness in
SNe Ia) are more uncertain.

\subsection{Nebular Phase}
\label{app:late}

The spectrum synthesis code for the nebular phases first calculates the energy
deposition by $\gamma$-rays and positrons produced in the decay of \Cofs. For a
given distribution of \Nifs, the energy deposition in each shell is evaluated in
a Monte Carlo scheme \citep{cappellaro97}. The deposited energy is spent in the
ionization and in the heating of thermal electrons. The latter is known to be
dominant in Fe-rich ejecta ($\sim 97$ per cent, \citealt{axelrod80}).

For given ionization fractions in the ejecta, the heating of the thermal
electrons ($\Gamma$)  is balanced with the radiative cooling (thermal balance):
\begin{equation}
\Gamma = \Gamma_{\rm line} + \Gamma_{\rm ff} + \Gamma_{\rm rec},
\end{equation}
where $\Gamma_{\rm line}$, $\Gamma_{\rm ff}$ and $\Gamma_{\rm rec}$ are the
cooling via lines, free-free emission, and recombination, respectively. Since 
the contribution from the latter two are smaller than the line cooling 
($\sim 0.03$ and $4$ per cent, respectively, 
\citealt{ruizlapuente92}), we only consider cooling by lines. The line cooling 
per unit volume by lines is given by
\begin{equation}
\Gamma_{\rm line} = \sum\limits_{k} n_{kj}
\sum\limits_{ij} A_{ji}h\nu_{ij}\beta_{ij},
\end{equation}
where $n_{kj}$ is the number density of the ion $k$ in the energy state $j$, 
and $A_{ji}$ and $\beta_{ij}$ represents for the Einstein A-coefficient and the
escape probability of the transition $j \rightarrow i$. From this thermal
balance, an electron temperature is evaluated in each shell.

The ionization fractions are determined balancing impact ionization
by non-thermal electrons and recombinations (ionization balance). Collisional
ionization by thermal electrons can be neglected \citep{ruizlapuente92}. The
ionization balance is solved for a given electron temperature. Using this new
ionization, the thermal balance is reconsidered, \ie the temperature and the
ionizations are solved iteratively. After the convergence, the emergent spectrum
is computed by integrating the contribution from each shell considering the
Doppler shift.

\section{Silicon Abundance at the Outer Layers}
\label{app:Si}

\begin{figure}
  \includegraphics[scale=0.70]{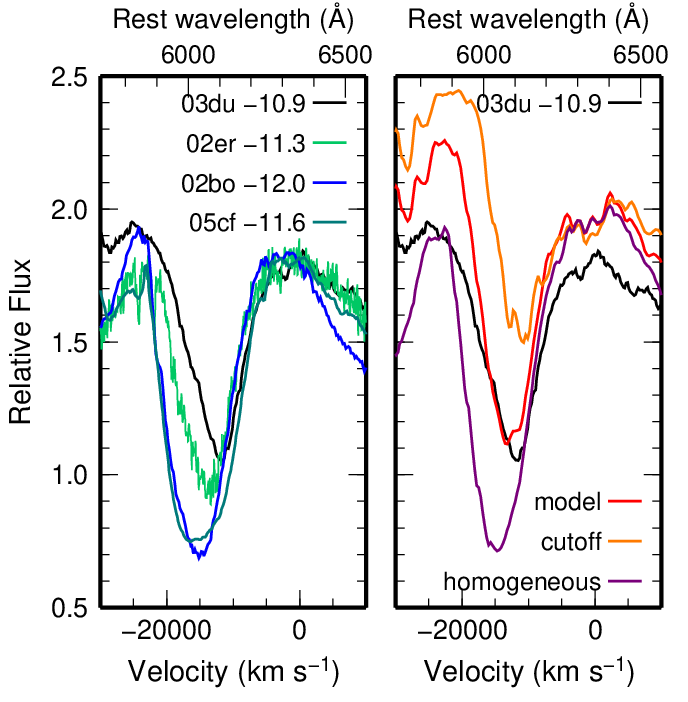} 
  \caption{The line profile of the \SiII\ $\lambda$6355 shown in the 
Doppler velocity measured from the rest wavelength of the Si {\sc II} line
($\lambda=6355$ \AA).
{\it Left:} The Si line profile in SNe 2003du (at $-10.9$ days from $B$ maximum, 
\citealt{stanishev07}), 
2002er ($-11.3$ days, \citealt{kotak05}), 2002bo ($-12.0$ days, \citealt{benetti04}), 
and 2005cf ($-11.6$ days, \citealt{garavini07})
at similar epochs.
These spectra are normalized to SN 2003du at $\lambda=6355$ \AA.
{\it Right:} The Si line profile in SN 2003du at $-10.9$ days from $B$ maximum
compared with three synthetic spectra.
Three model spectra are normalized to the best fit model at $\lambda=6355$ \AA.
The red line shows the best fit model using the abundance distribution
shown in Figs. \ref{fig:abunmass} and \ref{fig:abunvel}.
The orange line (``cutoff'') shows the model spectrum where 
the mass fraction of Si, $X$(Si), is set to be zero above 15000 \kms. 
The purple line (``homogeneous'') shows the model spectrum where 
the homogeneous Si abundance with $X$(Si)=0.3 is assumed at $v>10500$ \kms.
}
\label{fig:Si}
\end{figure}

The \SiII\ line profile in SN 2003du is rather narrow and it is similar to that
found in, \eg SNe 2002er \citep[Fig. \ref{fig:Si}, HVG,][]{kotak05}, 2003cg
\citep[LVG,][]{eliasrosa06}, 2005cg \citep[LVG,][]{quimby06}. 
This is in contrast to the broad
absorption seen in SNe 2002bo \citep[Fig. \ref{fig:Si}, HVG,][]{benetti04}, 
2002dj \citep[HVG,][]{pignata08}, 
2006X \citep[HVG,][]{wang0806X,yamanaka0906X}, or a possible boxy
profile in SNe 1990N \citep[LVG,][]{leibundgut9190N}, 
2001el \citep[LVG,][]{mattila05}, 2004dt
\citep[HVG,][]{altavilla07}, 
2005cf \citep[Fig. \ref{fig:Si}, LVG,][]{garavini07,wang0905cf}.
Here LVG and HVG mean low velocity gradient and high velocity gradient,
respectively. These are the classifications of Type Ia SNe 
based on the temporal evolution of the \SiII\ line velocity \citep{benetti05}.
Note that high velocity features of \CaII\ lines 
are commonly seen in both groups
at pre-maximum epochs \citep{mazzali05Ia}.

We use the spectrum at $-10.9$ days from $B$ maximum ($v_{\rm ph}=10500$ \kms) 
to constrain the distribution of Si in the outer layers because this spectrum is
reproduced better than the one at $-12.8$ days (Fig. \ref{fig:spec1}).
To reproduce the observed profile, a moderately mixed-out Si distribution is
required. The mass fraction of Si is $X$(Si)$\sim 0.3$ at $v=10500-15000$ \kms\
and it is suppressed at the outer layers, $X$(Si)$\sim 0.02-0.05$ at $v>15000$
\kms\ (see Figs. \ref{fig:abunmass} and \ref{fig:abunvel}). This model is shown
in Figures \ref{fig:spec1} and \ref{fig:Si} by the red line.

In the right panel of Figure \ref{fig:Si}, two other models are shown.  The
orange line ("cutoff") shows a model spectrum where the Si abundance has been
set to $X$(Si)=0 at $v>15000$ \kms\ as in the W7 model (Fig. \ref{fig:abunvel}).
In this model, the absorption of Si at $v>15000$ \kms\ is clearly too weak,
suggesting that some degree of burning took place in the outer layers
\citep{gerardy04}.

The purple line ("homogeneous") shows the model spectrum with a homogeneous Si
abundance with $X$(Si)=0.3 at $v>10500$ \kms. In this case, the absorption at a
high Doppler shift is too strong. This gives a similar profile to that in SNe
2002bo and 2005cf. In fact, such an extended Si distribution was derived for SN
2002bo in Paper I.

\section{Upper Limit for the Mass of Unburned Carbon}
\label{app:C}

\begin{figure}
  \includegraphics[scale=0.70]{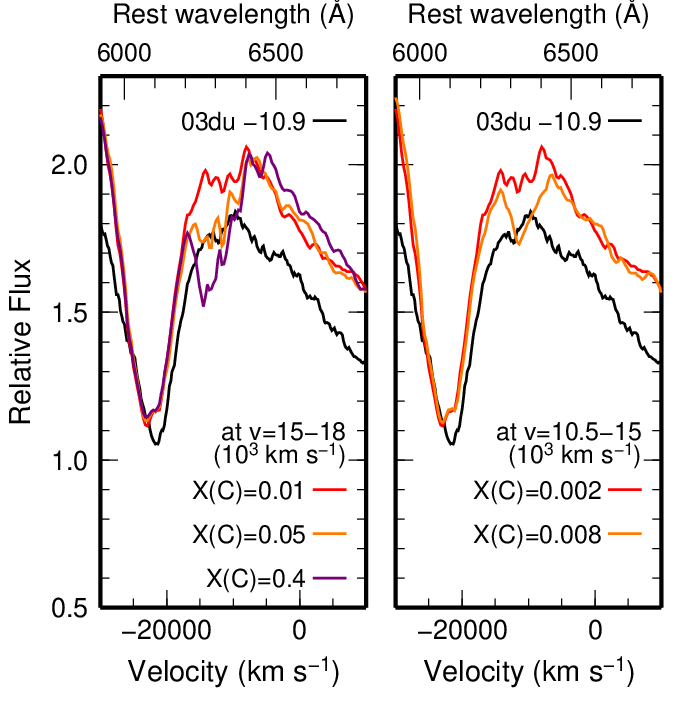} 
  \caption{
The line profile around the emission peak of the \SiII\ $\lambda$6355 line 
shown in the Doppler velocity measured from the \CII\ $\lambda$6578.
{\it Left:} The profile in SNe 2003du at $-10.9$ days from $B$ maximum
compared with the model spectra with  
$X$(C)=0.01 (red, best fit), 0.05 (orange), and 0.4 (purple) 
at $v=15000-18000$ \kms.
{\it Right:} The profile in SN 2003du at $-10.9$ days from $B$ maximum
compared with the model spectra with  
$X$(C)=0.002 (red, best fit) and 0.008 (orange) at $v=10500-15000$ \kms.
}
\label{fig:C}
\end{figure}

Lines of carbon are not clearly seen in the optical spectra of SN 2003du. A
marginal detection may be seen in correspondence of the emission peak of 
\SiII\ $\lambda$6355 \citep{stanishev07, tanaka08Ia}. The profile around
the emission peak is enlarged in Figure \ref{fig:C} and plotted versus Doppler
velocity measured from the \CII\ $\lambda$6578 line. We use the spectrum at
$-10.9$ days from $B$ maximum ($v_{\rm ph}=10500$ \kms) to estimate an upper
limit for unburned C since the emission peak of the Si line is reproduced more
nicely than in the spectrum at $-12.8$ days.

We first divide the ejecta at $v>10500$ \kms\ into three parts, with boundaries
at $v=15000$ and $18000$ \kms. At $v>18000$ \kms, no constraints can be obtained
on the C abundance because of the low density (weakness of the line) and the
blending with the strong \SiII\ line. Thus, we could set the upper limit of C as
$X$(C)=0.5 at $v>18000$ \kms. In practice, since there are other elements such 
as Si and Ca in this velocity space, 
the mass fraction is set to be $X$(C)=0.4 at $v>18000$ \kms\ in
the best fit model. Since the ejecta mass at $v>18000$ \kms\ is 
$2.0 \times 10^{-2} \Msun$, the upper limit of the C mass at $v>18000$ \kms\ is 
$1.0 \times 10^{-2} \Msun$.

The model spectrum with $X$(C)$=0.05$ at $v=15000-18000$ \kms\ shows a
noticeable \CII\ line at the emission peak of the \SiII\ line (left panel of
Fig. \ref{fig:C}). Since the ejecta mass at $v=15000-18000$ \kms\ is 
$6.0 \times 10^{-2} \Msun$, the upper limit of the C mass in this layer is 
$3 \times 10^{-3}
\Msun$, which is smaller than the C mass allowed in the outermost layers 
($1.0 \times 10^{-2} \Msun$ at $v>18000$ \kms).

At $v=10500-15000$ \kms, the constraint on the C abundance is even stronger. The
model spectrum with $X$(C) $=0.008$ gives a clear absorption trough of the \CII\
line (right panel of Fig. \ref{fig:C}). Thus, although the ejecta mass at 
$v=10500-15000$ \kms\ is large ($3.4 \times 10^{-1}\Msun$), the C mass at these
layers is less than $3 \times 10^{-3} \Msun$.

Thus, most of the C that is allowed must be located at $v>18000$ \kms.  Summing
up the upper limits to the C mass in different layers, we can safely conclude
that the mass of C ejected in SN 2003du is  $<1.6 \times 10^{-2} \Msun$ at $v >
10500$ \kms. This is less than in the deflagration models  ($4.86 \times 10^{-2}
\Msun$), where $X$(C)=0.5 at $v \gsim 15000$ \kms\ (see Figs. \ref{fig:abunmass}
and \ref{fig:abunvel}, and Table \ref{tab:mass}). Note that all estimates are
performed assuming the density structure of W7 model. If a delayed
detonation model was used, $X$(C) in the outermost layers would be smaller
\citep{tanaka08Ia}.

\end{document}